# On the Extended Hilbert Space in the theory of the multielectron atom

Alexey N. Hopersky, Alexey M. Nadolinsky *and* Rustam V. Koneev

*Rostov State Transport University, 344038, Rostov-on-Don, Russia*
*E-mail*: qedhop@mail.ru, amnrnd@mail.ru, koneev@gmail.com

**Abstract.** The construction of the Extended Hilbert Space (*EHS*) is presented in the form of a direct sum of the spaces of vectors of finite and infinite norms as the main space in the mathematical formalism of quantum mechanics of a multielectron atom. On the example of constructing the analytical structure of the probability amplitude $1s - 3p$ of photoexcitation of a neon atom, the implementation of the *EHS* – construct for solving the equations of the self-consistent Hartree-Fock field is given.

## Introduction

The methods of functional analysis [1–3], the theory of linear operators in Hilbert space [4–6], and the theory of generalized functions [7–10] make it possible to solve, in particular, one of the fundamental mathematical problems of quantum mechanics – the problem of constructing a complete orthonormalized set of wave functions of discrete and continuous spectra (space and excited) states of a multielectron atom. The crux of the problem is that the wave functions of continuous spectrum states have an infinite norm and thus cannot be functions of Hilbert space. As a result, finite norm functions from Hilbert space do not form the complete orthonormalized set needed to describe the quantum dynamics of an atom. In his monograph [11], Paul Dirac expressed the idea of searching for a "*more general than Hilbert's*" space of quantum mechanics using the concept of a generalized function $\delta(\varepsilon - \varepsilon') = \begin{cases} 0, \varepsilon \neq \varepsilon' \\ \infty, \varepsilon = \varepsilon' \end{cases}$ for the infinite norm of the wave function of a continuous spectrum ($\varepsilon \in [0; \infty)$), but he made no attempt to construct such a space. Of course, here and further we mean a functional ("distribution") $\delta(\varepsilon - \varepsilon') \rightarrow \langle \delta_{(\varepsilon)}, f \rangle = \int \delta(\varepsilon - \varepsilon') f(\varepsilon') d\varepsilon' = f(\varepsilon)$, where the *Dirac delta-function* is the kernel of the integral operator. In his monograph [12], Johann von Neumann made an attempt to mathematically solve the problem of taking into account the states of a continuous spectrum in the class of ordinary (locally integrable) functions of mathematical analysis, while remaining in Hilbert space and using the so-called "unit expansion" (an infinite family of projection operators [6]) for the Hermitian operator. Von Neumann's theory found no application to the specific problems of quantum mechanics of the multielectron atom and "turned out to be only one of those pretty faces that appear for a moment from the crowd, only to disappear in it forever" [13]. In the work of the authors [14], the idea of P. Dirac [11] is implemented through the construction of an Extended Hilbert Space. Mathematical formalism of the article [14] is expanded in our recent paper [15]. *In this article, we present the English version of the* work [15].

## Results

**Statement**. Let $\{|x_n\rangle\}$ be an infinite and countable ($n = 1, 2, \ldots, \infty$) system of linearly independent vectors of the finite norm: $\|x_n\| = \sqrt{\langle x_n | x_n \rangle} = 1$ and $\langle x_n | x_m \rangle \neq \delta_{nm}$ (Kronecker–Weierstrass symbol) at $n \neq m$. Let $\{|\varepsilon\rangle\}$ be a continuum ($\varepsilon \in [0; \infty)$) of vectors of infinite norm: $\langle \varepsilon | \varepsilon' \rangle = \delta(\varepsilon - \varepsilon')$ and $\langle \varepsilon | x_m \rangle \neq 0$. Then the system of vectors $\{|z_n\rangle\}$ and $\{|\varepsilon\rangle\}$ forms the complete orthonormalized basis of the infinite-dimensional Extended Hilbert Space as the direct sum of the *D*-Hilbert space of the finite norm vectors (the state of the *discrete* spectrum) and its extension (orthogonal complement) by the *C*-space of the vectors of the infinite norm (the state of the *continuous* spectrum):

$$EHS = D \oplus C, \quad \{|z_n\rangle\} \in D, \quad \{|\varepsilon\rangle\} \in C, \tag{1}$$

$$|z_n\rangle = \frac{|y_n\rangle}{\|y_n\|}, \quad |y_n\rangle = |\bar{x}_n\rangle + \sum_{i=1}^{n-1} \alpha_{in} |y_i\rangle, \tag{2}$$

$$|\bar{x}_n\rangle = a_n (|x_n\rangle - \int_0^\infty |\varepsilon\rangle\langle\varepsilon|x_n\rangle d\varepsilon), \quad \alpha_{in} = -\frac{\langle y_i | \bar{x}_n \rangle}{\|y_i\|^2}, \tag{3}$$

$$a_n = (1 - \int_0^\infty \langle \varepsilon | x_n \rangle^2 d\varepsilon)^{-1/2}, \tag{4}$$

$$\langle z_n | z_m \rangle = \delta_{nm}, \quad \langle \varepsilon | z_n \rangle = 0. \tag{5}$$

**Proof.** Let them be defined $\{|x_n\rangle\}$ (a system of linearly independent finite norm vectors) and $\{|\varepsilon\rangle\}$ (continuum of vectors of infinite norm). Let's redefine $\{|x_n\rangle\}$ as follows:

$$|x_n\rangle \to |\bar{x}_n\rangle = a_n(1-\hat{L})|x_n\rangle, \quad \langle\varepsilon|\bar{x}_n\rangle = 0. \tag{6}$$

In (6) $a_n$ is the normalization multiplier ($\|\bar{x}_n\| = 1$) and $\hat{L} = \int_0^\infty |\varepsilon\rangle\langle\varepsilon|d\varepsilon$ is the linear integral operator for the design of $D$-space to $C$-space. The vector system (6) becomes the starting point for the implementation of the *Gram-Schmidt orthogonalization* process [1,2]. After orthogonalization, we obtain an orthonormalized system of finite norm vectors $\{|z_n\rangle\}$. Complementing this system with a continuum $\{|\varepsilon\rangle\}$ -vectors of infinite norm, we get the direct sum of $D$- and $C$-spaces as *EHS*-space. At the same time, according to the (3), $C$ "*reflected*" in $D$ ($|z_n\rangle$ are defined through $|\varepsilon\rangle$). Due to the completeness of the set $|z_n\rangle$- and $\{|\varepsilon\rangle\}$- of vectors the closure condition is met:

$$\hat{L} + \hat{P} = \delta(r-r'), \tag{7}$$

where the $C$-space to $D$- space projection operator is defined $\hat{P} = \sum_{n=1}^{\infty} |z_n\rangle\langle z_n|$. Acting as an operator (7) on an arbitrary vector $|\psi\rangle \in EHS$ and integrating software $r'$, get presentation $|\psi\rangle$ – vectors in an orthonormalized basis $|\psi\rangle = \int_0^\infty |\varepsilon\rangle\langle\varepsilon|\psi\rangle d\varepsilon + \sum_{n=1}^{\infty} |z_n\rangle\langle z_n|\psi\rangle$ and its norm (Parseval's generalized equality)

$\|\psi\| = \left(\int_0^\infty \langle\varepsilon|\psi\rangle^2 d\varepsilon + \sum_{n=1}^{\infty} \langle z_n|\psi\rangle^2\right)^{1/2}$, where the convergence of the improper integral and the series is assumed. <u>Statement has been proven.</u>

As an example of the implementation of the *EHS*-space for solving the Hartree–Fock equations, we find the matrix element ($M$) of the single-electron $\hat{r}$-operator of the radiation transition as the amplitude of the $1s \to 3p$ probability of photoexcitation of the neon atom (Ne; the charge of the atom nucleus Z = 10; the configuration and term of the ground state $[0] = 1s^2 2s^2 2p^6[^1S_0]$): $\hbar\omega + [0] \to 1s(2s^2 2p^6)3p(^1P_1)$, where $\hbar$ is Planck's constant and $\omega$ is circular frequency of the photon being absorbed. According to the **Statement**,

$$M = \eta\left(M_0 - \int_0^\infty M(\varepsilon)\langle\varepsilon p_+|3p_+\rangle d\varepsilon\right), \quad M, M_0, M(\varepsilon) = \langle 1s_0\|\hat{r}\|3\tilde{p}_+, 3p_+, \varepsilon p_+\rangle,$$

$$\langle 1s_0\|\hat{r}\|3p_+\rangle = N_{1s}\left(\langle 1s_0|\hat{r}|3p_+\rangle - \frac{\langle 1s_0|\hat{r}|2p_+\rangle\langle 2p_0|3p_+\rangle}{\langle 2p_0|2p_+\rangle}\right), \quad N_{1s} = \langle 1s_0|1s_+\rangle\langle 2s_0|2s_+\rangle^2\langle 2p_0|2p_+\rangle^6, \tag{8}$$

$$\eta = \left(1 - \int_0^\infty \langle\varepsilon p_+|3p_+\rangle^2 d\varepsilon\right)^{-1/2}, \quad \langle 3\tilde{p}_+|2p_+\rangle = 0, \quad \langle 3p_+|2p_+\rangle \neq 0, \quad \langle 3\tilde{p}_+|\varepsilon p_+\rangle = 0, \quad \langle 3p_+|\varepsilon p_+\rangle \neq 0.$$

Structures (8) arise from the implementation of the methods of the theory of non-orthogonal orbitals [16]. The indices "0" and "+" correspond to the radial parts of the wave functions of the electrons obtained by the solution of the equations of the self-consistent Hartree–Fock field [17] for the configurations of the ground ($[0]$) and excited ($[1s_+^2 2s_+^2 2p_+^6 3p_+]$) states of the atom Ne. According to the **Statement**, we have: $|x_1\rangle = |2p_+\rangle$, $|x_2\rangle = |3p_+\rangle$ and $|z_2\rangle = |3\tilde{p}_+\rangle$. The numerical solution of the Hartree–Fock equations gives a noticeable (~ 7 %) difference in probabilities $1s \to 3p$ photoexcitation of the atom Ne: $\frac{P_0}{P} \sim \left(\frac{M_0}{M}\right)^2 = 1.07$, where $P_0(P)$ is probability without taking into account (taking into account) the design of the *EHS*- space.

**Comments**

**1.** B. Szőkefalvi-Nagy introduced the concept of "Extended Hilbert Space" (see **Appendix 1** to the monograph[3]) as the direct sum of several Hilbert spaces. In this case, in order to describe the states of the *continuous* spectrum, it is inevitable to refer either to the von Neumann's theory [12] (see **Introduction**) or, for example, to the method of this article.

**2.** If the source vector systems $\{|x_n\rangle\}$ and $\{|\varepsilon\rangle\}$ orthonormalized [$\langle x_n|x_m\rangle=\delta_{nm}$, $\|x_n\|=1$, $\langle\varepsilon|x_n\rangle=0$, $\langle\varepsilon|\varepsilon'\rangle=\delta(\varepsilon-\varepsilon')$], then the process of Gram-Schmidt orthogonalization loses its meaning and is defined *EHS*-space (without "reflecting" $C$ in $\overline{D}$):

$$EHS = \overline{D}\oplus C, \ \{|x_n\rangle\}\in\overline{D}, \ \{|\varepsilon\rangle\}\in C. \tag{9}$$

Space (9) arises, for example, in the description of the states of a single-electron hydrogen atom ($1s[^2S_{1/2}]$), where it is possible (in contrast to the Hartree–Fock equations) to *analytically* solve the stationary Schrödinger equations: $\hat{H}|x_n\rangle=x_n|x_n\rangle$, $\hat{H}|\varepsilon\rangle=\varepsilon|\varepsilon\rangle$, $\hat{H}$ is the atom operator of Hamilton.

**3.** In article [18], it was found that in the Hartree–Fock approximation for the integral of the overlapping radial parts of wave functions of continuous spectra, we have:

$$\langle\varepsilon|\varepsilon'\rangle=\delta(\varepsilon-\varepsilon')+\mathcal{P}\left(\frac{f(\varepsilon,\varepsilon')}{\varepsilon-\varepsilon'}\right), \tag{10}$$

where $\mathcal{P}$ is the symbol of the principal meaning in Cauchy's sense and $f(\varepsilon,\varepsilon')\neq 0$ at $\varepsilon\neq\varepsilon'$. Thus, the *nonlocality* of the *exchange* potential in the Hartree–Fock equation *violates* the requirement of normalizability of wave functions of a continuous spectrum by the "pure" $\delta$-Dirac function. Formally, mathematically, result (10) has nothing to do with the **Statement**, but indicates the need for an *analytical* modification of the Hartree-Fock approximation, which is widely used in the literature.

**4.** According to (3), the peculiarity of the direct sum (1) is the fact that *D*-space *becomes* Hilbert *through* the *C*-space of vectors of an infinite norm ($\langle x_n|x_m\rangle\neq\delta_{nm}\rightarrow\langle z_n|z_m\rangle=\delta_{nm}$; "from-itself-to-itself through the other-itself" [19]).

**Disclosures**

The authors declare no conflicts of interest.

**Data availability**

Data underlying the results presented may be obtained from the authors upon reason-able request.